\documentclass{ws-ijmpe}
\usepackage[compress]{cite}
\usepackage{isotope}
\usepackage[symbol]{footmisc}
\usepackage{hyperref}
\hypersetup{
    colorlinks=true, 
    linktoc=all,     
    linkcolor=blue,  
    citecolor=blue
}

\def\openone{\leavevmode\hbox{\small1\kern-3.8pt\normalsize1}}
\newcommand{\gsim}{\lower.7ex\hbox{$\;\stackrel{\textstyle>}{\sim}\;$}}
\newcommand{\lsim}{\lower.7ex\hbox{$\;\stackrel{\textstyle<}{\sim}\;$}}

\begin{document}

\markboth{O. Benhar \& L. Tonetto}{Neutrino emission from neutron star matter}

\catchline{}{}{}{}{}

\title{Neutrino emission from neutron star matter}

\author{\footnotesize Omar Benhar$^\dagger$ and Lucas Tonetto$^\ddagger$}

\address{INFN, Sezione di Roma \\ 
Dipartimento di Fisica, Sapienza Universit\`a di Roma \\
 I-00185 Roma, Italy\\
$^\dagger$omar.benhar@roma1.infn.it  \\
$^\ddagger$lucas.tonetto@uniroma1.it }

\maketitle

\begin{history}
\received{Day Month Year}
\revised{Day Month Year}
\end{history}

\begin{abstract}
The temperature of a  newly formed neutron star is believed to be as high as $10^{11}$~K, corresponding to a thermal 
energy of about $10$ MeV. After a time $t \sim 50 \ {\rm s}$, the neutrino mean free path in nuclear matter exceeds 
the typical star radius, R~$\sim$~10 Km, and neutrino emission becomes the dominant mechanism of 
energy loss, eventually 
bringing the temperature down to $\sim10^{8}$~K. Neutrinos also play a critical role in determining the composition 
of matter in the star interior, consisting primarily of a charge-neutral mixture of neutrons, protons and leptons 
in $\beta$-equilibrium. This article provides an introduction to the weak interactions of nucleons in nuclear matter, 
as well as a concise review of the neutrino emission reactions taking place in the neutron star core. The approximations 
involved in the standard theoretical treatment of thermal and dynamical effects are analysed in the light of the recent 
progress of the field, and the prospects for future developments are outlined.
\end{abstract}


\ccode{PACS numbers:25.30.Pt,13.15.+g,26.60.-c,24.10.Cn}


\section{Introduction}
\label{intro}

The development of a consistent theoretical framework for the description of neutrino emission and propagation in dense 
matter is a prerequisite for large-scale simulations of a number of astrophysical processes, including supernova explosions, 
neutron star cooling, and binary neutron star mergers; see, e.g., Refs.~\cite{refId0,Mezzacappa:2020oyq}. 

Neutrinos are a primary factor in the thermal evolution of neutron stars, and are critical for the determination 
of matter composition in the star interior~\cite{Prakash:1996xs,Yakovlev:2004iq}.  
A prominent role is played by the neutrino emission processes taking place in the neutron star core, which is 
believed to consist predominantly of a charge-neutral fluid of neutrons, protons and leptons in weak equilibrium.

Although the fundamental theories of strong and weak interactions predict 
the appearance of different forms of matter\textemdash such as strange baryonic matter or quark matter\textemdash in the high-density
 limit, the validity of the description in terms of nucleons is strongly supported by Bayesian inference analyses of recent astronomical data~\cite{Brandes:2023hma,Brandes:2024wpq}. 
The results of these studies indicate that a phase transition is, in fact, unlikely to occur even in the most massive neutron star observed so far~\cite{Romani:2022jhd}, the central density of which 
exceeds the density of atomic nuclei by a factor of $\sim 5$. As recently pointed out by the authors of Ref.~\cite{Brandes:2023hma}, independent experimental evidence reinforcing this conclusion is provided by the 
emergence of $y$-scaling in electron-nucleus scattering data; see Ref.\cite{Benhar:2023qvr} and references therein.

Neutrinos are also emitted from the low-density matter comprising the neutron star crust through a variety of reasction mechanisms.
However, a discussion of these processes\textemdash which are known to be important, and must be taken into account in numerical 
studies of neutron star cooling\textemdash is outside the scope of our work. The interested reader is
referred to the exhaustive reviews of Yakovlev {\it et al.}~\cite{Yakovlev} and Chamel and Haensel~\cite{Chamel:2008}.

It is important to realise that, in spite of the fact that neutrino-nucleon interactions in free space can be understood within 
the framework of the standard model of particle physics, the description of weak interactions in nuclear matter involves 
a host of challenging issues, originating mainly from the complexity of nuclear structure and dynamics. Additional difficulties 
arise from the treatment of thermal effects in protoneutron stars and in the remnants of neutron star mergers, the 
temperatures of which lie in the tens of MeV range.

The simplest reactions leading to neutrino and antineutrino emission from nuclear 
matter, referred to as Urca processes,  are neutron $\beta$-decay and the capture of the associated charged lepton by a proton.
Early analyses of these processes\textemdash involving a number of simplifying assumptions expected to be applicable 
in the low-temperature regime\textemdash revealed that the requirement of momentum conservation leads the emergence of a lower 
bound for the proton density. This result has important implications because, depending on the nuclear matter equation of state, 
the appearance of the threshold may prevent activation of the Urca processes in most neutron stars.

In the absence of Urca processes, the dominant neutrino and antineutrino emission reactions, dubbed modified Urca, proceed 
through a mechanism in which momentum conservation is enabled by the presence of an additional nucleon, acting as a spectator. 
As a consequence, the calculation of the corresponding rates requires a model of the nucleon-nucleon interaction.
In their groundbreaking work of the late 1970s~\cite{Friman:1979ecl}, Friman and Maxwell employed a number of low-temperature 
approximations and a simplified model of nuclear dynamics, based on the one-pion-exchange potential.
The results of the analysis of Ref.~\cite{Friman:1979ecl}\textemdash which was pursued further  and 
generalised by the authors of Refs.~\cite{Yakovlev_AA,Yakovlev}\textemdash provided the baseline for later 
theoretical studies of the modified Urca process.

A significantly improved description of nuclear dynamics\textemdash involving the use of a realistic 
Hamiltonian and the $G$-matrix formalism\textemdash was employed by Shternin~{\it et al.} to obtain the
modified Urca reaction rates reported in Ref.~\cite{Shternin:2018dcn}. This work can be seen as a pioneering attempt to study neutrino emission from nuclear matter using a realistic nuclear Hamoltonian and the formalism of nuclear many-body theory.

Recently, significant efforts have been made to go beyond the approximations involved in the work of Friman and 
Maxwell~\cite{Suleiman:2023bdf,Alford:2024xfb,Sedrakian:2024uma,Lucas:Thesis,tobepublished}. These studies, 
featuring both a detailed treatment of thermal effects and a more realistic description of nuclear dynamics,  
were strongly motivated by the advent of multimessenger astronomy, and the ensuing unprecedented progress of neutron 
star observations; see, e.g., Ref.~\cite{Benhar:2024qcw}.

Ideally, neutrino emission from neutron stars should be described within a unified framework, in which 
the impact of temperature and nuclear dynamics on all relevant nuclear matter properties\textemdash from the 
equation of state, to the nucleon effective masses and chemical potentials\textemdash are consistently taken 
into account. Such an approach has been pursued by the authors of Refs.~\cite{Lucas:Thesis,tobepublished}, who 
employed an effective Hamiltonian derived from state-of-the-art phenomenological potentials~\cite{BL:2017}, and 
the formalism for the treatment of hot nuclear matter discussed in Refs.~\cite{Benhar_2022,PhysRevD.106.103020}.  

This article is intended to provide a concise description of the main processes leading to neutrino emission from 
nuclear matter, as well as a review of the results of selected theoretical approaches employed for their study.
The remainder of the manuscript is organised as follows. After a brief introduction to the weak interactions of isolated 
nucleons, Sect.~\ref{Nweak} analyses the corresponding processes in nuclear matter, with an emphasis on the impact of 
nuclear dynamics beyond the mean-field approximation. Section~\ref{mechanisms} is devoted to the discussion of the Urca 
and modified Urca reactions, and to a description of the theoretical models employed to obtain the corresponding neutrino 
emission rates.  Finally, the concluding section summarises the present status of the field and lays down the prospects 
of future developments.

\section{Weak interactions in nuclear matter} 
\label{Nweak}

In the energy regime relevant to neutron star physics, weak interactions involving leptons and baryons can be accurately described replacing the 
exchange of  the gauge bosons $W^{\pm}$ and $Z_0$\textemdash the masses of which are $\sim$80 and $\sim$90 GeV, respectively\textemdash  with a contact 
four-fermion interaction. This tenet lies at the basis of the celebrated Fermi theory of neutron $\beta$-decay.

\subsection{Nucleon weak interactions in free space} 
\label{freenucleon}

The Fermi Lagrangian describing charged-current (CC) and neutral-current (NC) nucleon weak 
interactions\textemdash associated with $W^{\pm}$ and $Z_0$ exchange, respectively\textemdash
can be written as a current-current product in the form
\begin{align}
\label{lagrangian}
\mathcal{L}_F(x) = - \frac{G_w}{\sqrt{2}} \ {\ell}_\alpha(x) {j}^\alpha(x)   \ \ \ \ \ ,  \ \ \ \ \ \ G_w = \left\{
\begin{array}{ll}
G_F & \ \ (NC) \\
G_F \cos \theta_C & \ \ (CC)
\end{array}
\right. \ ,
\end{align}
with $G_F = 1.166 \times 10^{-5}  \ {\rm GeV}^{-2}$ and $\theta_C = 13.02^\circ$ being the Fermi coupling constant and the Cabibbo mixing
angle, respectively. The explicit expressions of the lepton currents, $\ell_\alpha(x)$, are
\begin{align}
\label{currents:rel}
\ell_\alpha(x) = \left\{
\begin{array}{ll} 
{\overline \psi}_{\ell}(x) \gamma_\alpha(1-\gamma_5)\psi_\nu(x) & \ \ (CC) \\
\overline{\psi}_{\nu}(x) \gamma_\alpha(1-\gamma_5)\psi_\nu(x) & \ \ (NC)
\end{array} 
\right.  \ \ , 
\end{align}
with $\psi_\nu$ and $\psi_\ell$ being the Dirac fields describing a neutrino and the associated charged 
lepton. The nucleon currents can be conveniently written in the form 
\begin{align}
\label{nucleon:rel}
j^\alpha(x) = \left\{
\begin{array}{ll}
 \overline{\Psi}_{N}(x) \gamma^\alpha(g_V- g_A \gamma_5)\tau^\pm \Psi_{N}(x) & \ \ (CC) \\
   \overline{\Psi}_{N}(x)  \frac{1}{2} \gamma^\alpha(c_V-c_A \gamma_5)\Psi_{N}(x) & \ \ (NC)
\end{array}
\right. \  ,
\end{align}
where $g_V$ and  $g_A$ ($c_V$ and $c_A$) are the CC (NC) vector and axial-vector neutrino-nucleon coupling constants. The nucleon 
field $\Psi_N$ is an isospin doublet, the upper and lower components of which are the proton and neutron fields $\psi_p$, and $\psi_n$, 
while  $\tau^\pm $ denotes the isospin raising and lowering operators, associated with $W^\mp$ exchange, respectively.

Neutral current interactions in nuclear matter have been discussed extensively in studies  
of neutrino-pair bremsstrahlung, as well as in investigations of reactions involving more exotic particles, such as axions; see, e.g., Refs.~\cite{Raffelt:1996wa,Hanhart:2000ae}. This paper, on the other hand, is limited to a discussion of
charged current interactions, which are known to provide the dominant contribution to neutrino and antineutrino 
emission from neutron stars.
 
In the non relativistic limit, routinely employed to describe the nucleon, the upper line of the right-hand side of Eq.~\eqref{nucleon:rel} is  replaced by 

\begin{align}
\label{currents:NR}
{j^\pm}^\alpha \equiv ({j^\pm}^0,{\bf j}^\pm)  \ \ , \ {\rm with} \  \ \left\{
\begin{array}{ll}
 {j^\pm}^0 = g_V~{\rm e}^{i {\bf q} \cdot {\bf x} }~\tau^\pm  \\~\\
{\bf j}^\pm  = g_A~{\rm e}^{i {\bf q} \cdot {\bf x} }~\boldsymbol{\sigma}~\tau^\pm \\
\end{array}
\right. \  ,
\end{align} 
where ${\boldsymbol \sigma} \equiv(\sigma_1,\sigma_2,\sigma_3)$, with the $\sigma_i$'s being Pauli matrices, describes the nucleon spin, while ${\bf q}$ denotes the momentum transfer to the nucleon.

\subsection{Weak interactions in matter and nuclear dynamics} 
\label{matter}

Theoretical studies  of CC weak interaction processes in nuclear matter involve the evaluation of transition amplitudes that can be written in the form 
\begin{align}
\label{nuclear:ampl}
M^\alpha_{0F} = \langle F |~{J^+}^ \alpha~| 0 \rangle \ , 
\end{align}
with
\begin{align}
\label{nuclear:current}
J^{+^ \alpha} = \sum_{i=1}^A~j_i^{+^\alpha}  \ , 
\end{align}
where $j_i^{+^\alpha}$ is the current operator acting on the $i$-th nucleon and $A$ is the number of nucleons\footnote{In nuclear matter, both $A$ and the normalisation volume $V$ tend to infinity, 
with the baryon  number density $\varrho_B = A/V$ remaining finite.}. For any proton and neutron densities $\varrho_p$ and $\varrho_n$, the initial and final states of Eq.~\eqref{nuclear:ampl} are solutions of the many-body Schr\"odinger equations 
\begin{align}
\label{schroedinger}
{\rm H} |0\rangle = E_0 |0\rangle \ \ \ \ \ , \ \ \ \ \ {\rm H} |F\rangle = E_F |F\rangle \ ,
\end{align}
with 
\begin{align}
\label{Hamiltonian}
{\rm H} = \sum_{i=1}^A \frac{{\bf k}_i^2}{2 m_N} + \sum_{j>i=1}^A v_{ij} + \sum_{k>j>i=1}^A V_{ijk} \ . 
\end{align}
Here, $m$ is the nucleon mass, ${\bf k}_i$ denotes the momentum of the $i$-th nucleon, and  the potentials $v_{ij}$ and $V_{ijk}$ describe nucleon-nucleon (NN) and irreducible 
three-nucleon (NNN) interactions, respectively.
Advanced models of the Hamiltonian provide a quantitative account of the observed properties of the two- and three-nucleon systems\textemdash which are  calculable to 
high accuracy\textemdash and are well suited to explain the equilibrium properties of isospin-symmetric matter inferred from nuclear systematics.
It should be pointed out that the basic assumption underlying the treatment of hot matter within nuclear many-body theory is that 
at low-to-moderate temperatures\textemdash typically $T \ll m_\pi$, $m_\pi \approx 140$ MeV being the pion mass\textemdash nuclear 
dynamics, as described by the Hamiltonian of Eq.~\eqref{Hamiltonian}, is largely unaffected by thermal effects.

Owing to the strong spin-isospin dependence of the potentials appearing in Eq.~\eqref{Hamiltonian}, the solution of the Schr\"odinger equation using non perturbative methods\textemdash  notably those based on Quantum Monte Carlo 
techniques~\cite{QMC}\textemdash is out of reach of the present computing capabilities for nuclei having $A \geq 12$. On the other hand, the use of standard perturbation theory is hampered by the strongly repulsive nature of
short-range NN interactions, which makes the matrix elements of $v_{ij}$ too large for a perturbative expansion to converge. 
As a consequence, the evaluation of the nuclear transition amplitude of Eq.~\eqref{nuclear:ampl} unavoidably requires the introduction of simplifying assumptions.   

\subsubsection{The mean-field approximation} 
\label{MFA}

The mean-field approximation (MFA)\textemdash underlying the nuclear shell model\textemdash is based on the tenet that nucleons can be treated as independent particles 
subject to an average potential $U_N$, with $N=n,p$, similar to the quasiparticles of Landau's theory of normal Fermi liquids~\cite{Landau}. Within this scheme, the 
lowest-energy state of nuclear matter at temperature $T$, baryon number density $\varrho_B$ and proton fraction $x_p=\varrho_p/\varrho_B$ can be  written in terms of 
eigenstates of the mean-field Hamiltonian belonging to the eigenvalues
\begin{align}
\label{SP:hamiltonian}
e_N(k) = \frac{ {\bf k^2} }{ 2m } + {U}_N({k}) \ ,  
\end{align}
with $k = |{\bf k}|$, distributed according to the Fermi-Dirac function
\begin{align}
n_N(k,T)  = \Big\{ 1 + {\rm e}^{[ e_N(k) - \mu_N]/T } \Big\}^{-1} \ . 
\label{FD:distribution}
\end{align}
The above equation describes the probability of finding a nucleon of species $N$ in the state specified by momentum ${\bf k}$, the 
energy of which, $e_N(k)$, is given by  Eq.~\eqref{SP:hamiltonian}.
The chemical potential $\mu_N$ is determined by the constraint
\begin{align}
\label{def:chempot}
\frac{2}{V} \sum_{\bf k} n_N({k},T) = \varrho_N \ , 
\end{align}
where $V$ is the normalisation volume, and the factor 2 accounts for the spin degeneracy of momentum eigenstates. It follows that, in general, 
$\mu_N$ depends on $T$ and the particle number density $\varrho_N$. 

Because the nuclear weak current defined by Eq.~\eqref{nuclear:current} is the sum of  single-nucleon operators,  in order for the transition 
amplitude $M^\alpha_{0F}$ not to vanish, the MFA initial and final states can only differ by one orbital. 
Therefore, $|F\rangle$ is obtained from the ground state by replacing a neutron of momentum ${\bf k}_n$, distributed according to $n_n({k}_n,T)$,  with a proton, the momentum of which, ${\bf k}_p$, follows the distribution $1 - n_p({k}_p,T)$. The energies and momenta of the states $|F \rangle$ and $|0 \rangle$ are trivially related through $E_F = E_0 - e_n(k_n) + e_p(k_p)$, and ${\bf P}_F = {\bf P}_0- {\bf k}_n + {\bf k}_p$, respectively.

Note that, in the $T \to 0$ limit, $n_N({k},T)  \to \theta [ \mu_N - e_N({k}) ]$, with $\theta(x)$ being the Heaviside step function, and $\mu_N \to  e_{F_N}$. Here, 
the Fermi energy  is defined as $e_{F_N} = e_N(k_{F_N})$, with the Fermi momentum ${k_{F_N}}$ being related to the corresponding number density through 
 $k_{F_N} =  (3 \pi^2 \varrho_N)^{1/3}$. The resulting $T=0$ ground state consistes of neutrons and protons occupying all energy levels corresponding
to momenta such that $k_n < k_{F_n}$ and  $k_p < k_{F_p}$, while $|F\rangle$ is a one-particle\textendash one-hole state, in which a neutron 
with $k_n <  k_{F_n}$ is replaced by a proton with $k_p >  k_{F_p}$.
 
\subsubsection{Nuclear dynamics beyond the mean field and short-range correlations} 
\label{correlations}

While being able to explain a number of important properties of atomic nuclei, the MFA is inherently incomplete, in that it fails to take into account the effects of 
short-range
correlations (SRC) among the nucleons, originating mainly from strongly repulsive nuclear forces.
The most prominent signature of SRC is the appearance of high-momentum nucleons in the aftermath of virtual NN scattering processes, associated with a 
suppression of the probability of finding two nucleons within the range of the repulsive interactions. In Nuclear Many-Body Theory (NMBT), 
this {\it screening} effect is effectively accounted for through a {\it renormalisation} of the NN potential of Eq.~\eqref{Hamiltonian}, leading to the 
determination of a density-dependent {\it effective interaction}, suitable  to carry out perturbative calculations of nuclear matter properties.

Unambiguous evidence of correlation effects in nuclei\textemdash first observed in a series of pioneering studies of electron-induced proton 
knockout reactions, concisely reviewed in Ref.~\cite{Benhar_NPN}\textemdash has been provided by a wealth of electron-nucleus scattering data, collected 
over many decades using a 
variety of targets; see, e.g., Refs.~\cite{RevModPhys.65.817,Arrington_SRC}.

Two main approaches\textemdash outlined in, e.g., Ref.~\cite{BF:NM}\textemdash have been followed to tame the non perturbative nature of the NN interaction.
The first one is based on the replacement of the bare $v_{ij}$ with the well-behaved operator describing NN 
scattering in the nuclear medium. This effective interaction, referred to as  $G$-matrix, can be used in perturbation theory in conjunction with the basis of eigenstates of the non interacting system. 
In the alternative approach, on the other hand, the NN potential $v_{ij}$  is kept unchanged, while the Fermi gas basis is replaced with a complete set of states embodying the screening effect originating from SRC. 

The {\it correlated} ground state, $|0 \rangle$,  is obtained from the transformation 
\begin{align}
\label{def:F}
|  0 \rangle =  \frac{ {\mathcal F}~| 0 ) }{ \sqrt{ (0 | {\mathcal F}^\dagger {\mathcal F} |0) } }\ , 
\end{align}
where $|0)$ denotes the Fermi gas ground state and
\begin{align}
\label{def:fcorr}
{\mathcal  F} = \Big\{ {\mathcal S} \prod_{j>i=1}^A~f_{ij} \Big\} \ , 
\end{align}
with the operator structure of the NN {\it correlation functions} $f_{ij}$ reflecting the spin-isospin dependence of 
the potential $v_{ij}$. Note that,  in view of the fact that  $[f_{ij},f_{jk}] \neq 0$, the product appearing in the right-hand side of the above equation needs to 
be symmetrised through the action of the operator ${\mathcal S}$.

The radial dependence of the correlation functions is obtained by solving a set of Euler-Lagrange equations resulting from functional minimisation of the 
expectation value of the Hamiltonian of Eq.~\eqref{Hamiltonian} in the correlated ground state of nuclear matter at $T=0$. 
The choice of determining the correlation functions in  cold matter is supported by the results of numerical calculations at $T\neq0$, showing that the temperature dependence of  $f_{ij}$ 
remains negligibly small for temperatures as high as  $\sim20$ MeV.  

In the wake of the pioneering work of Cowell and Pandharipande~\cite{cowell2003}, the correlated ground state defined by  Eqs.~\eqref{def:F} and~\eqref{def:fcorr} has been employed to define a density-dependent effective interaction, denoted 
${\widetilde v}_{ij}$, through the relation
\begin{align}
\label{def:veff}
\langle 0 | {\rm H} | 0 \rangle 
= ( 0 |~\Big\{ \sum_{i=1}^A \frac{ {\bf k}_i^2 }{ 2m } + \sum_{j>i=1}^A {\widetilde v}_{ij}\Big\}~| 0 ) \ ,  
\end{align}
with ${\rm H}$ being the nuclear Hamiltonian of Eq.~\eqref{Hamiltonian}. 
For any fixed nucleon density $\varrho_B$, ${\widetilde v}_{ij}$ is obtained by applying the cluster expansion technique, thoroughly described in Ref.~\cite{CLARK197989}, 
which amounts to rewrite the left-hand side of the above equation as a sum of contributions arising from subsystems, or clusters, involving an increasing 
number of nucleons; a detailed discussion of the determination of ${\widetilde v}_{ij}$ from a nuclear Hamiltonian involving both NN and NNN potentials can be found in Refs.~\cite{BL:2017,Benhar_2022}. 

The resulting effective interaction is independent of temperature by construction, and designed to reproduce the ground-state energies of cold isospin-symmetric nuclear 
matter (SNM) and pure neutron matter (PNM) obtained using advanced computational approaches, such as the Auxiliary Field Diffusion Monte 
Carlo (AFDMC) method or the variational scheme known as Fermi Hyper-Netted Chain/Single Operator Chain (FHNC/SOC) summation. 
It should be pointed out that the formalism based on the effective interaction ${\widetilde v}_{ij}$ 
allows for a consistent treatment of equilibrium and dynamical properties of nuclear matter within a unified framework. A notable early application was the calculation of the 
shear viscosity and thermal conductivity of PNM described in Refs.~\cite{benharvalli2007,Benhar:2009xm}.
In addition, ${\widetilde v}_{ij}$ can be used to derive the mean field $U_N$ from the nuclear Hamiltonian of Eq.~\eqref{Hamiltonian}. 
In the Hartree-Fock approximation, the resulting expression turns out to be
\begin{align}
\label{HF:potential}
U_N({k}) = \sum_{{\bf k}^\prime} ( {\bf k} {\bf k}^\prime |~\sum_{j>i=1}^A {\widetilde v}_{ij}~| {\bf k} {\bf k}^\prime \rangle_A~n_N( {k}^\prime, T ) \ ,
\end{align}
where $n_N$ is the Fermi-Dirac distribution of Eq.~\eqref{FD:distribution}, and $| {\bf k} {\bf k}^\prime \rangle_A = | {\bf k} {\bf k}^\prime \rangle - | {\bf k}^\prime {\bf k} \rangle$ denotes the antisymmetric state of two non-interacting nucleons.

\subsection{Effective weak current}
\label{eff:operator}

The correlation operator ${\mathcal F}$ of Eq.~\eqref{def:fcorr} 
is also employed to obtain the complete set of states referred to as correlated basis functions, or CBF, defined as
\begin{align}
| n \rangle =  \frac{ {\mathcal F}~| n ) }{ \sqrt{ ( n | {\mathcal F}^\dagger {\mathcal F} |n) } }\ , 
\end{align}
where $|n)$ denotes an excited state of the non-interacting Fermi gas.
The set $\{| n \rangle\}$ is suitable to perform calculations of the nuclear transition amplitudes of Eq.~\eqref{nuclear:ampl} including the effects
of SRC; for a concise review of the application of the CBF formalism to nuclear matter, see, e.g., Ref.~\cite{BF:NM}.

It should be kept in mind that, within the independent-particle model, the structure of the nuclear matter eigenstates appearing in  Eq.~\eqref{nuclear:ampl} is dictated only
by translation invariance and Fermi-Dirac statistics. As a consequence, the transition amplitude $M^\alpha_{0F}$ obtained from  the Fermi gas model is left unchanged 
by the inclusion of nuclear dynamics at mean-field level.
On the other hand, the results of a number of theoretical studies provide convincing evidence that SRC lead to a sizeable suppression 
of  weak interactions in nuclear matter; see, e.g., Refs.~\cite{cowell2006,benharfarina2009,Lovato:2012ux,lovatoetal2014}. 

These findings can be understood considering that the same mechanism accounted for by renormalisation of the nuclear Hamiltonian\textemdash that is, the appearance of high-momentum nucleons and the ensuing depletion of the Fermi gas eigenstates\textemdash  also leads to 
a quenching of the transition amplitude of Eq.\eqref{nuclear:ampl}, which can likewise be described in terms of a    
{\it renormalised current operator} and Fermi gas states. 

Within the approach based on the CBF formalism, the effective weak current $\widetilde{j}_i^+$ is defined  
by the equation
\begin{align}
\label{def:jeff}
\langle {F} | J^{+^ \alpha} | 0 \rangle = ( {F} |~\sum_{i=1}^{A} \widetilde{j}_i^{+^\alpha}|  {0} )  \ ,  
\end{align} 
to be compared to Eq.~\eqref{def:veff}. The derivation of the effective current $\widetilde{j}_i^+$ from the above equation, based on the cluster expansion technique, is
discussed in Ref.~\cite{lovatoetal2014}.

\begin{figure}[th]
\centerline{\includegraphics[width=7.25cm]{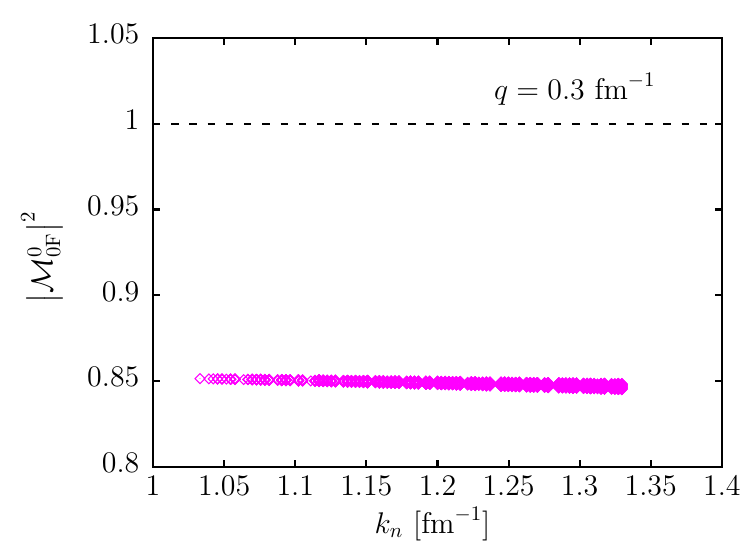}}
\caption{\scriptsize Squared transition matrix element of the $\alpha=0$ component of 
the effective weak current, defined by Eq.~\eqref{def:jeff},  in  cold SNM at equilibrium density. 
The results, corresponding to fixed momentum transfer $q=0.3 \ {\rm fm}^{-1}$, are displayed as a function of the neutron momentum. For comparison, the 
dashed horizontal line shows the predictions of the independent-particle model. Adapted from Ref.~\cite{benharfarina2009}.}
\label{amplitude}
\end{figure}

As an example, Fig.~\ref{amplitude} shows the squared transition matrix element of the $\alpha=0$ component of 
the effective weak current, defined according to Eq.~\eqref{def:jeff},  in  isospin-symmetric nuclear matter at equilibrium density, $\varrho_0 = 0.16 \ {\rm fm}^{-3}$, 
and $T~=~0$. The calculation was performed by the authors of Ref.~\cite{benharfarina2009}, using the CBF formalism and correlation functions 
obtained from a realistic nuclear Hamiltonian including NN and NNN interactions. The results, corresponding to fixed momentum transfer $q = |{\bf k}_p - {\bf k}_n| = 0.3 \ {\rm fm}^{-1}$, show 
that the inclusion of SRC leads to a significant departure from the predictions of the independent-particle model\textemdash represented by the dashed 
horizontal line\textemdash largely independent of the magnitude of the neutron momentum. 

\section{Neutrino emission mechanisms}
\label{mechanisms}

This paper is limited to a discussion of neutrino and antineutrino emission from charge-neutral and $\beta$-stable matter 
consisting of neutrons, protons and electrons, the densities of which are denoted $\varrho_n$, $\varrho_p$, and $\varrho_e$, respectively. In such a system, referred to as $npe$ matter, electrons are treated as ultra 
relativistic non interacting particles, while interactions among the nucleons are described 
within the non relativistic framework outlined above. Neutrinos and antineutrinos are assumed to stream freely 
through matter, with their densities and chemical potentials being therefore vanishing. 
The extension of the formalism to take into account the appearance of muons\textemdash which is energetically 
favoured when the electron chemical potential exceeds the muon rest mass\textemdash does not involve any conceptual issues. 

In principle, the microscopic description of nuclear matter 
should take into account possible deviations from the Fermi liquid behaviour associatd with the appearance of superfluid and superconducting phases~\cite{Sedrakian:2018ydt}, as well as with the onset of pion condensation~\cite{Takatsuka:1993}.
However, although the emergence of these non standard forms of nuclear matter may affect neutrino emission and neutron star cooling\textemdash see, e.g., Refs.~\cite{Pethick:RMP,Muto:1992,Tsuruta:2002}\textemdash their discussion is beyond the scope of the present work.

\subsection{Direct Urca processes}
\label{Urca}

The simplest weak interaction processes leading to $\nu$ and ${\overline \nu}$ emission from $npe$ matter are neutron $\beta$-decay  and the inverse electron capture reaction 
\begin{align}
\label{durca}
        n   \rightarrow p + e + {\overline \nu} \ \ \ \ \ , \ \ \ \ p + e  \rightarrow n + \nu \ .
\end{align}
The above reactions\textemdash first discussed by Gamow and Schoenberg in the 1940s~\cite{gamow:Urca}, and famously named 
Urca processes~\cite{gamow}\textemdash 
provide the most effective cooling mechanism  of massive neutron stars; see, e.g., Ref.\cite{Pethick:RMP}. 
In view of the discussion of the more complex neutrino emission mechanism to be discussed in Sect.~\ref{modified:Urca},  
hereafter the processes~\eqref{durca} will be referred to as direct Urca, or dUrca.

At equilibrium, neutron decay and electron capture contribute equally to the neutrino emissivity\textemdash defined as the total $\nu$ and ${\overline \nu}$ energy emitted per unit time and 
volume\textemdash  and the chemical potentials of the degenerate fermions satisfy the condition
\begin{align}
\label{chemical:equilibrium}
\mu_n = \mu_p + \mu_e \  .
\end{align}
The solution of equation~\eqref{chemical:equilibrium}  yields the equilibrium value of the proton fraction $x_p=\varrho_p/\varrho_B$, which fully 
determines matter composition at fixed $T$ and $\varrho_B$. 

\begin{center} 
\begin{figure}[h!] 
\centering
\includegraphics[width=3.25cm]{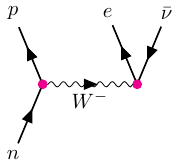}
\caption{Diagrammatic representation of ${\overline \nu}$ emission through the dUrca process in $npe$ matter.}
\label{dUrca}
\end{figure}
\end{center}

For illustration, let us consider ${\overline \nu}$ emission associated with the elementary process illustrated by the diagram of Fig.~\ref{dUrca}. 
The corresponding transition rate can be obtained using Feynman diagram techniques, keeping in mind that the external nucleon lines and the vertices 
{\it do not} represent non interacting fermions and their weak interactions in free space. The resulting expression can be cast in the form
\begin{align}
\label{W:Urca}
W_{U} = \frac{  (G_F \cos \theta)^2}{2} \frac{1}{4 E_e E_\nu} L_{\alpha \beta} S^{\alpha \beta} \ , 
\end{align}
where the tensor  $L^{\alpha \beta}$ can be readily written in terms of  the lepton four-momenta,   $k_e~\equiv~(E_e,{\bf k_e})$ and $k_\nu \equiv(E_\nu,{\bf k_\nu})$, as
\begin{align}
L_{\alpha \beta} =  8 \left[ {k_e}_\alpha {k_\nu}_\beta + {k_e}_\beta {k_\nu}_\alpha - g_{\alpha\beta}(k_e k_\nu)  + i \epsilon_{\alpha\beta\rho\sigma} k_e^\rho k_\nu^\sigma \right] \ ,
\label{leptens}
\end{align}
with $g_{\alpha\beta} = {\rm diag}(1,-1,-1,-1)$ and $\epsilon_{\alpha\beta\rho\sigma}$ being the metric tensor and the four-dimensional
Levi-Civita tensor, respectively.  

The expression of  $S^{\alpha \beta}$, involving the transition matrix elements of the nucleon weak current $M^{\alpha}_{0F}$ defined by Eq.~\eqref{nuclear:ampl}, reads
\begin{align}
\label{nuctens}
S^{\alpha \beta}~=~\sum_F~M^{\alpha  \ \dagger}_{0F}  M^{\beta}_{0F}~(2 \pi)^4 \delta^{(4)}( P_0 - P_F - k_e - k_\nu ) \ ,  
\end{align}
where $P_0 \equiv (E_0,{\bf P}_0)$ and  $P_F \equiv (E_F,{\bf P}_F)$ denote the four-momenta of the nuclear matter initial and final states. Within 
the independent particle model underlying the MFA, $S^{\alpha \beta}$ turns out to be diagonal, and the transition rate of the dUrca 
process\textemdash a pedagogical derivation of which can be found in, e.g., Ref~\cite{Benhar:LNP}\textemdash  reduces to
\begin{align}
\label{urcarate:MFA}
W_{U} & = 2 (G_F \cos \theta_C)^2 \Big[ g_V^2 (1 + \cos\theta)  + g_A^2 ( 3 - \cos\theta) \Big] \\ 
 \nonumber
&   \ \ \ \ \ \ \ \ \times (2 \pi)^{4}\delta\left( E_n  - E_p - E_e - E_\nu \right)\delta^{(3)}\left( {\bf k}_n - {\bf k}_p - {\bf k}_e - {\bf k}_\nu \right) \ , 
\end{align} 
where $k_N \equiv(E_N,{\bf k}_N)$ denotes a nucleon four-momenta, $E_n = e_n(k_n)$, $E_p = e_p(k_p)$, and $\theta$ is the angle between the 
lepton momenta ${\bf k}_e$ and ${\bf k}_\nu$. 

In the low-temperature regime relevant to neutron stars, all fermions participating in the dUrca processes are strongly degenerate, their momenta being close to 
the corresponding Fermi momenta. Under these conditions, the widely used Fermi surface approximation (FSA), which amounts to setting $|{\bf k}_i| =  k_{F_i}$
with $i = n, p , e$, is  expected to be reliably applicable. Assuming, in addition, that the neutrino momentum\textemdash the typical magnitude of which is $|{\bf k}_\nu| \sim T$\textemdash can be neglected with respect to the nucleon and electron Fermi momenta, the equation expressing momentum conservation can be written in the form 
\begin{align}
\label{mom:cons}
{\bf k}_n = {\bf k}_p + {\bf k}_e \ , 
\end{align}
which entails the inequality
\begin{align}
\label{Urca:triangle}
k_{F_p} \geq k_{F_n} - k_{F_e} \ .
\end{align}
By combining the above result with the constraints associated with baryon number conservation and charge neutrality, 
implying  $\varrho_p = \varrho_B - \varrho_n =  \varrho_e$, one obtains the condition $x_p = \varrho_p/\varrho_B \geq x_{\rm thr} =  1/9$, to be satisfied 
by the proton fraction $x_p$ in order for the dUrca mechanism to be active.  

\begin{figure}[th]
\centerline{\includegraphics[width=6.25cm]{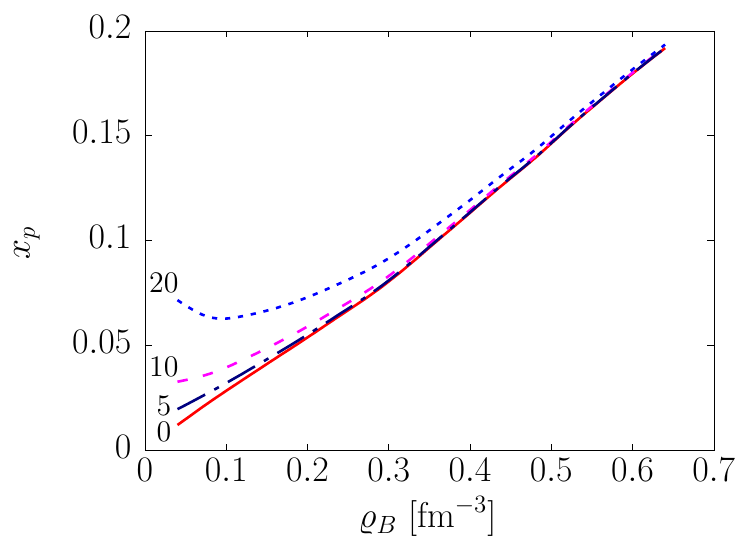}\includegraphics[width=6.25cm]{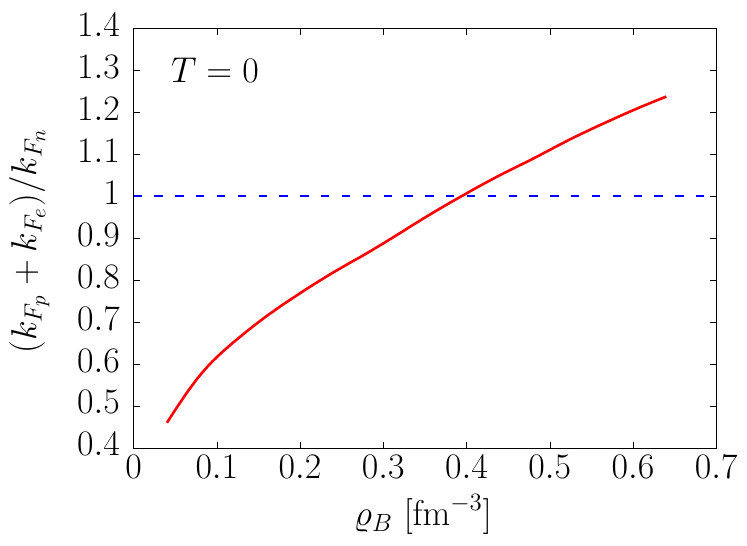}}
\caption{\scriptsize Left panel: density dependence of the proton fraction in $npe$ matter at temperature $T=$0, 5, 10, and 20 MeV. 
Adapted from Ref.~\cite{PhysRevD.106.103020}.
Right panel: density dependence of the ratio $(k_{F_p} + k_{F_e})/k_{F_n}$. The horizontal line indicates the threshold for activation
of the dUrca mechanism at $T = 0$. Reprinted from Ref.~\cite{Lucas:Thesis}. \copyright~Lucas Tonetto, 2024. All rights  reserved.}
\label{proton_fraction}
\end{figure}

Given a model of the nuclear mean field $U_N$, which determines the chemical potentials through Eqs.~\eqref{SP:hamiltonian}\textendash\eqref{def:chempot}, 
the equilibrium value of the proton fraction $x_p$ at temperature $T$ and baryon density $\varrho_B$ is obtained from the solution of Eq.~\eqref{chemical:equilibrium}. 
As an example, the left panel of Fig.~\ref{proton_fraction} shows the density dependence of the proton fraction of charge-neutral, $\beta$-stable matter at temperatures in the range $T=0-20$ MeV 
reported in Ref.~\cite{PhysRevD.106.103020}. The effective interaction employed in these 
studies was derived by the authors of Refs.~\cite{BL:2017,Benhar_2022} using the CBF formalism and a phenomenological nuclear 
Hamiltonian comprising the Argonne $v_{6}^\prime$  NN potential~\cite{V6P} and 
  and the Urbana IX NNN potential~\cite{UIX_2,UIX}. Thermal effects turn out to be significant, and lead to  
  a departure from the monotonic behaviour typical of cold matter for $T\geq 20$ MeV. 
  
  In the low-temperature regime, in which the proton fraction increases 
  with $\varrho_B$, the appearance of the threshold $x_{\rm thr}$ entails the occurrence of a corresponding threshold density $\varrho_{\rm thr}$, 
  determined by the condition $(k_{F_p} + k_{F_e})/k_{F_n} = 1$. The results displayed in the right panel of Fig.~\ref{proton_fraction}\textemdash obtained from 
  the zero-temperature proton fraction of the left panel\textemdash show that $\varrho_{\rm thr} \approx 0.4 \ {\rm fm}^{-3}$. This relatively low value, exceeding nuclear saturation density by a factor of about 2.5, suggests that 
  the dUrca process may, in fact, be allowed in neutron stars with the canonical mass $M= 1.4~M_\odot$. 
  
  
It should be emphasised that the activation threshold of the Urca processes only appears in the limit of zero-temperature, which implies the onset 
of the strong degeneracy regime underlying the FSA. 
While laking a consistent interpretation at $T >0$, however, the threshold 
obtained from the above procedure turns out to be useful as a baseline for the analysis of the $T$-dependence of $npe$ matter composition; see, 
e.g., Ref.~\cite{Benhar:2023mgk}.  

\subsection{Emissivity of the dUrca processes}
\label{direct:Urca}

The neutrino emissivity associated with the dUrca processes is obtained from integration of the product between the reaction rate $W_U$, defined by 
Eqs.~\eqref{W:Urca}-\eqref{nuctens}, and a factor describing the total phase space available to the participating particles, shaped by the Fermi-Dirac distributions. The resulting expression is
\begin{align}
\label{Q:Urca}
Q_{U} = 2 \, \int \  W_{U} \, E_\nu \ n_n (1-n_p) (1-n_e)~\frac{d^3 k_n}{( 2 \pi)^3} \frac{d^3 k_p}{( 2 \pi)^3} \frac{d^3 k_e}{( 2 \pi)^3}  \frac{d^3 k_\nu}{( 2 \pi)^3}\, ,
\end{align}
where the factor 2 is needed to include the contributions of both $\nu$ and ${\overline \nu}$. 



\subsubsection{Low-temperature approximation}
\label{lowT}

In the low-temperature regime, the calculation of $Q_U$ can be greatly simplified by neglecting ${\bf k}_\nu$
in the momentum-conserving $\delta$-function, which allows to readily perform the $\cos \theta$ integration. In addition, the remaining integrations are routinely carried out
by applying the so-called phase space decomposition~\cite{Yakovlev}, which amounts to substituting
\begin{align}
\label{PSD}
\frac{d^3 k_n}{( 2 \pi)^3} \frac{d^3 k_p}{( 2 \pi)^3} \frac{d^3 k_e}{ ( 2 \pi)^3} \longrightarrow \prod_{i=n,p,e} {k_F}_i m^\ast_i dE_i d \Omega_i \ , 
\end{align} 
as prescribed by the FSA. Here, $\Omega_i$ is the solid angle specifying the direction of the momentum ${\bf k}_i$, and the fermion effective masses 
are defined as
\begin{align}
\label{effmass}
\frac{1}{m^\star_i} = \frac{1}{k_{F_i}}  \left( \frac{d E_i}{ d |{\bf k}_i |}\right)_{ |{\bf k}_i| = {k_F}_i } \ .
\end{align}
Within the MFA, in which the reaction rate is given by Eq.~\eqref{urcarate:MFA}, the resulting expression of the emissivity can be cast in the form~\cite{Yakovlev}
\begin{align}
\label{Urca:FSA}
       Q_D \propto  \, G_F^2 \cos^2{\theta_c}\,(1+3g_A^2) \, m^*_n\,m^*_p\,m^*_e  \, T^6 \, \Theta_{npe} \ , 
\end{align}
with
\begin{align}
\label{theta:npe}
         \Theta_{npe}  =
        \begin{cases}
                1 & \text{if $ k_{F_n} \leq k_{F_p} + k_{F_e} $} \\
                0 & \text{otherwise} 
        \end{cases} \ ,
\end{align}
showing a specific power-law dependence on temperature\textemdash dictated by the number of degenerate fermions participating in the process\textemdash and 
the threshold behaviour as a function of proton fraction discussed above. Because both these features are determined by the use of the FSA, the occurrence of deviations from the predictions of 
Eq.~\eqref{Urca:FSA}\textemdash which may originate from the failure of the low-temperature approximation as well as from effects of  nuclear 
dynamics beyond the MFA\textemdash needs to be carefully analysed.

\subsubsection{Thermal and correlation effects}
\label{Improved:Urca}

\begin{figure}[ht]
\centerline{\includegraphics[width=6.00cm]{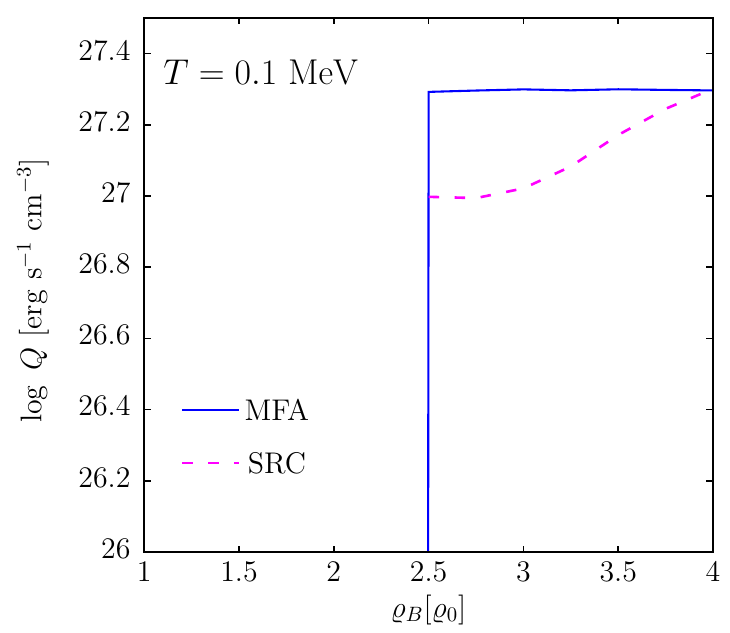}}
\caption{\scriptsize Density dependence of the neutrino emissivity associated with the Urca processes in $npe$ matter at $T=0.1 \ \mathrm{MeV}$, computed using the FSA and the phase space decomposition. 
The solid and dashed lines represent the results of obtained within the MFA ad including the effects of SRC, respectively. Note the  
activation threshold of the Urca processes at $\varrho = 2.5~\varrho_0$. Reprinted from Ref.~\cite{Lucas:Thesis}. \copyright~Lucas Tonetto, 2024. All rights  reserved.}
\label{Urca_SRC}
\end{figure}

Figure \ref{Urca_SRC} shows the density dependence of the emissivity of the Urca processes in $npe$ matter at T~=~0.1 MeV, evaluated 
within the approach  based on the FSA and the phase space decomposition described above; the effective masses and chemical potentials 
of the degenerate fermions involved in the calculation of $Q_U$\textemdash derived from  the CBF effective interaction defined by Eq.~\eqref{def:veff}\textemdash are thoroughly discussed in Ref.~\cite{PhysRevD.106.103020}. 

The MFA predictions, represented by the solid line, are compared to the results obtained by replacing the free-space nucleon weak current, $j^+_i$, with the renormalised operator embodying the effects of SRC ${\widetilde j}^+_i$, defined by Eq.~\eqref{def:jeff}. The details of the derivation of ${\widetilde j}^+_i$ using the CBF 
formalism and the cluster expansion technique  can be found in  Ref.~\cite{Lucas:Thesis}. Is is apparent that the inclusion of nuclear dynamics beyond the MFA, while not affecting the 
occurrence of the threshold at $\varrho_{\rm thr} = 0.4~{\rm fm}^{-3}$\textemdash consistent with the $T=0$ result shown in the right panel of Fig.~\ref{proton_fraction}\textemdash leads to significant modifications of the density dependence of $Q_U$. The emissivity turns out to be reduced by as much as  50 \% at $\varrho \approx  \varrho_{\rm thr}$, and monotonically increases with  $\varrho$. The MFA value is approached at 
$\varrho \approx 0.64~{\rm fm}^{-3}$, corresponding to four times nuclear matter saturation density.

\begin{figure}[th]
\centerline{\includegraphics[width=6.25cm]{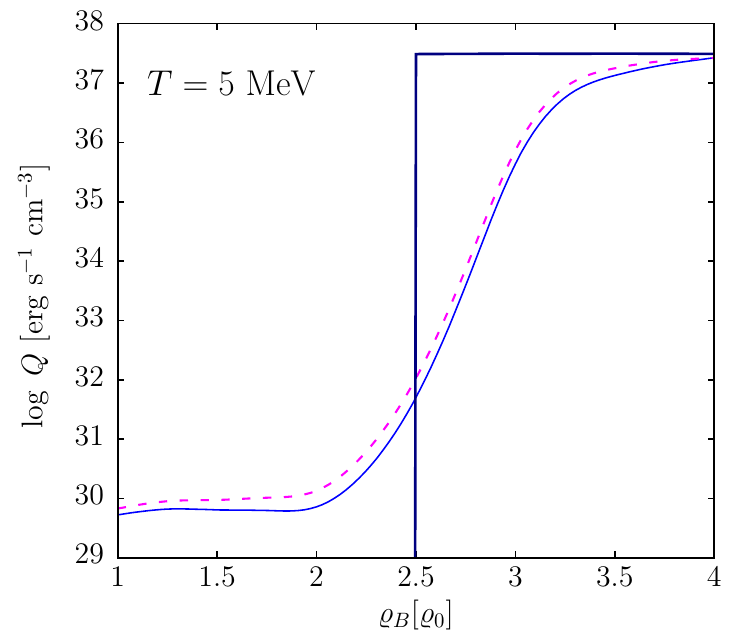}\includegraphics[width=6.25cm]{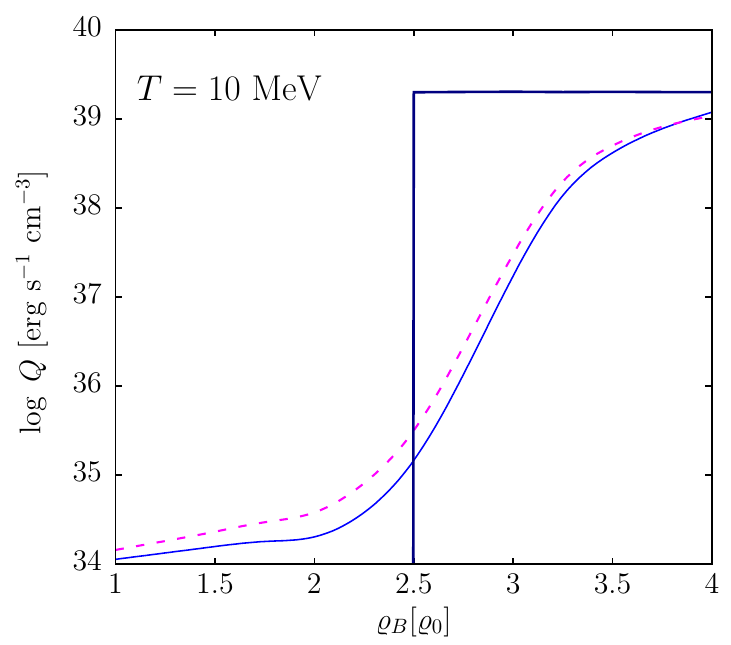}}
\caption{\scriptsize Emissivity of the dUrca processes at $T$ = 5 and 10 MeV, obtained from Eq.~\eqref{Q:Urca} without using the FSA, and carrying out the full phase space integration.
The solid and dashed lines correspond to calculations performed with and without inclusion of the SRC, respectively.
For comparison, the thick solid lines at $\varrho > 2.5~\varrho_0$ show the results obtained using the low-temperature approximations 
described in the text. Reprinted from Ref.~\cite{Lucas:Thesis}. \copyright~Lucas Tonetto, 2024. All rights  reserved.}
\label{Urca_full}
\end{figure}

In order to shed light on  the breakdown of the low-temperature approximation, the emissivities of the dUrca processes  
at temperature $T=$ 5 and 10 MeV have been also evaluated lifting the simplifying assumptions implied in the FSA, and using Monte Carlo techniques to carry out the full phase space integrations involved in Eq.~\eqref{Q:Urca}~\cite{Lucas:Thesis}. The results of these calculations, displayed in the two panels of Fig.~\ref{Urca_full}, clearly show that, owing to the thermal broadening of the Fermi surface, the sharp threshold observed in Fig~\ref{Urca_SRC} is washed out already  at $T=5$ MeV. The emissivity turns out to be a smooth 
function of density over the whole range $1 \leq (\varrho_B/\varrho_0) \leq 4$, and approaches the predictions of the low-temperature approximation\textemdash shown by the thick solid line\textemdash with increasing $\varrho_B$. Note that at $T=10$ MeV the asymptotic value is reached at larger density. 

\begin{center} 
\begin{figure}[h!] 
\centering
\includegraphics[width=6.40cm]{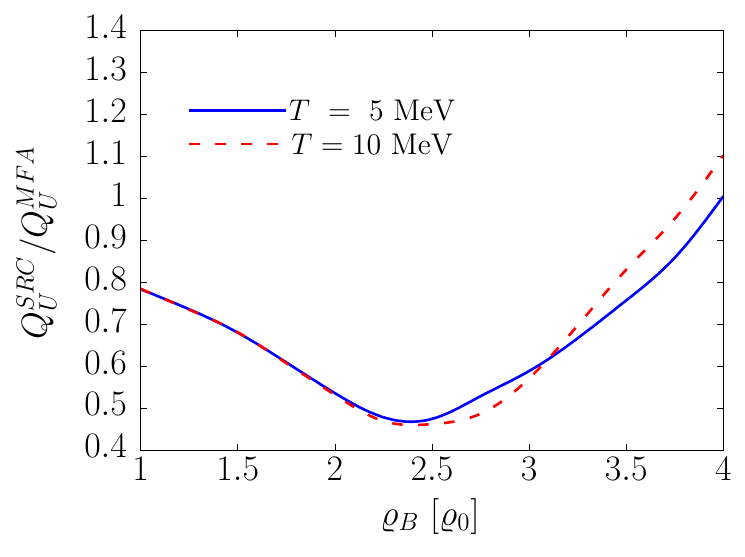}
\caption{Density dependence of the ratio between the emissivities of the dUrca processes computed with and without inclusion of SRC at temperature $T$ = 5 and 10 MeV. Adapted from Ref.~\cite{Lucas:Thesis}.}
\label{Qratio}
\end{figure}
\end{center}

The solid and dashed lines of Fig.~\ref{Urca_full}, corresponding to the emissivities obtained with and without inclusion of SRC\textemdash which 
will be denoted $Q_U^{SRC}$ and $Q_U^{MFA}$, respectively\textemdash exhibit largely similar behaviours. The impact of SRC can be best identified considering the density dependence of the ratio $Q_U^{SRC}/Q_U^{MFA}$, displayed 
in Fig.~\ref{Qratio} for both $T$= 5 and 10 MeV. 

All in all, the picture emerging from Figs.~\ref{Urca_full} and~\ref{Qratio} indicates that, regardless of the inclusion of correlation effects,  
releasing  the low-temperature approximation leads to significant modifications of the emissivity at all densities. 
Note that the $\sim50$\% quenching of the ratio $Q_U^{SRC}/Q_U^{MFA}$ occurring at $\varrho_B \approx\ 2.5~\varrho_0$, is reminiscent of the 
$T= 0.1$ MeV result of Fig.~\ref{Urca_SRC}.  The differences between the curves corresponding to different temperatures, observed Fig.~\ref{Qratio} for large values of $\varrho_B$, 
can be understood considering the interplay between thermal and density-dependent dynamical effects on equilibrium and non-equilibrium properties of nuclear matter, discussed in Ref.~\cite{Benhar:2023mgk}.

\subsection{Modified Urca processes}
\label{modified:Urca}

\begin{figure}[h!] 
\centering
\includegraphics[width=8.0cm]{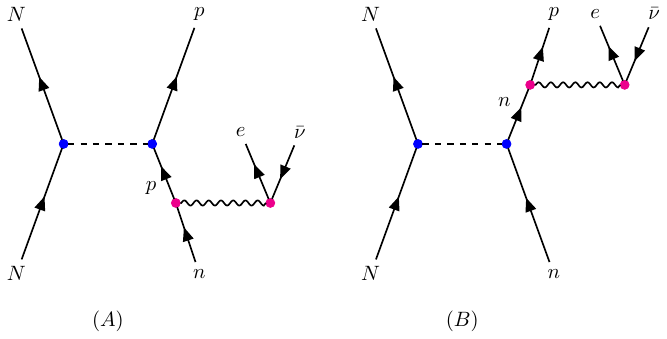}
\caption{Diagrammatic representation of ${\overline \nu}$ emission through modified Urca processes in $npe$ matter.
Both diagrams involve a combination of neutron $\beta$-decay and a NN interaction, depicted by the dashed line. The label $N=n,p$ corresponds  to the neutron and proton branch, respectively.}
\label{mUrca}
\end{figure}

The modified Urca, or mUrca, processes
\begin{align}
n + N \to p + e + {\overline \nu} + N \ \ \ \ \ , \ \ \ \ \ p + e + N \to n + \nu + N \ , 
\end{align}
where $N=n,p$ labels the neutron and proton branch, respectively, 
involve combinations of $\nu$ and ${\overline \nu}$ emission through the dUrca reactions~\eqref{durca} and a 
NN collision; two examples of mUrca processes are illustrated by the diagrams of Fig.~\ref{mUrca}.
Note that, owing to the presence of the NN interaction, the MFA is inherently inadequate to describe the mUrca mechanism. 

It should be reminded that the dashed lines of Fig.~\eqref{mUrca} {\it do not} represent the NN potential $v_{ij}$ appearing in the nuclear Hamiltonian of Eq.\eqref{Hamiltonian}, which is not amenable to treatment by perturbation theory. In their seminal study of neutrino emissivities associated with 
mUrca processes~\cite{Friman:1979ecl}, Friman and Maxwell circumvented this problem using a simplified model of $v_{ij}$, comprising the 
long-range one-pion-exchange (OPE) interaction, supplemented by a short-range interaction parametrised according to the scheme of 
Landau's theory of Fermi liquids~\cite{Landau}. 

A comparison between Figs.~\ref{dUrca} and \ref{mUrca} shows that, in addition to the dashed line representing the NN interaction, the diagrams 
depicting mUrca processes involve an internal fermion line, representing the nucleon propagator $G_N$. The authors of Ref.~\cite{Friman:1979ecl}, 
whose results were obtained within the FSA described above and neglecting the momentum of the outgoing antineutrino, employed the 
approximate expression $G_N~\approx~1/E_e$.

\begin{figure}[th]
\centerline{\includegraphics[width=7.50cm]{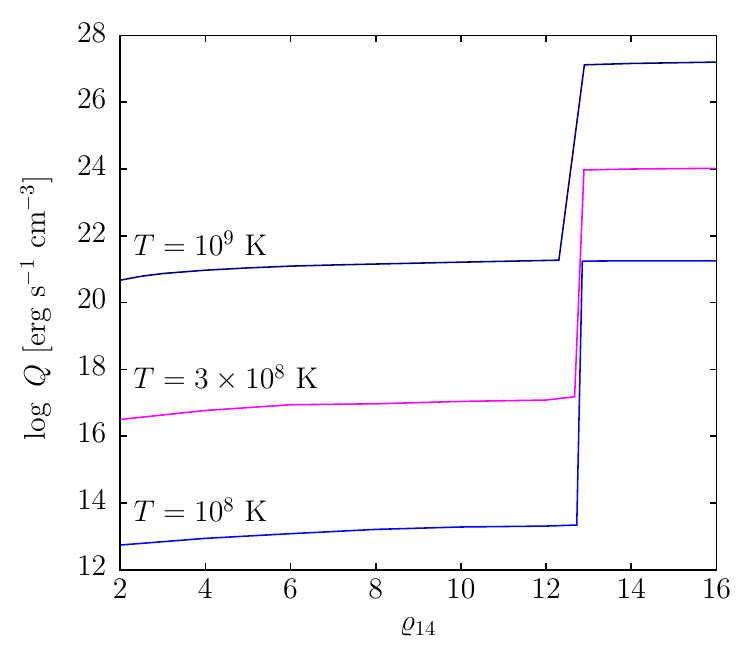}}
\caption{\scriptsize Total neutrino emissivity in $npe$ matter at $T = 10^8, \ 3\times 10^8$, and $10^9$~K, displayed as
a function of matter density measured in units of $10^{14} \ {\rm g  \ cm}^{-3}$. Note that in these units the 
saturation density of cold nuclear matter is $\varrho_0 \approx 2.57$. The results have been obtained by Yakovlev {\it et al.} using the low-$T$ approximation
discussed in the text and the formalism based on the work of Friman and Maxwell~\cite{Friman:1979ecl}. Adapted from Ref.~\cite{Yakovlev}.}
\label{Q:Yakovlev}
\end{figure}

More detailed analyses of neutrino emission based on the conceptual framework underlying the work of Friman and Maxwell have been carried out by 
Yakovlev~{\it et~al.}~\cite{Yakovlev_AA,Yakovlev}. The results of these studies, illustrated in Fig.~\ref{Q:Yakovlev} for temperatures in the range 
$10^8 - 10^9$ K\textemdash  corresponding to thermal energies $8.62 - 86.2$ keV\textemdash clearly show the threshold behaviour as a function of baryon density, 
expected in the $T \to 0$ limit. Note, however, that at densities close to the dUrca threshold, small deviations from the sharp discontinuity described by the step function of Eq.~\eqref{theta:npe} are approximately taken into account by an exponential suppression of the emissivity. The effect of this correction is clearly visible at $T=10^9$ K. 

The dominant high-density component originates from the onset of the dUrca mechanism discussed in Sec.~\ref{direct:Urca}, while the much smaller sub-threshold contribution is associated with mUrca processes. The threshold values of matter density\textemdash $\varrho  \gtrsim 1.2\times10^{15}$ g/cm$^3$, corresponding to baryon number densities  $\varrho_B \gtrsim 0.72 \ {\rm fm}^{-3}$, or 4.5~$\varrho_0$\textemdash  suggest that, according to the model employed by the authors, the dUrca process is only  likely to be active in massive neutron stars, the central density of which exceeds the saturation density of SNM by a factor $4-5$. 

An improved treatment of the neutrino emissivity originating from mUrca processes was proposed by Shternin {\it et al.} in Ref.~\cite{Shternin:2018dcn}. In this work, the renormalised NN interaction was derived from a realistic nuclear 
Hamiltonian\textemdash comprising the Argonne $v_{18}$~\cite{AV18} NN potential and the UIX NNN 
potential~\cite{UIX_2,UIX}\textemdash using the $G$-matrix formalism. 
The nucleon propagator was written in the quasiparticle form 
\begin{align}
\label{QP:propagator}
G^{\rm MFA}_N(k,E) = \frac{1}{E - e_N(k)} \ , 
\end{align}
with the nucleon energy of Eq.~\eqref{SP:hamiltonian} computed in Brueckner Hartee-Fock approximation using 
Eqs.~\eqref{SP:hamiltonian} and~\eqref{HF:potential}. The results reported in Ref.~\cite{Shternin:2018dcn}\textemdash obtained by applying the FSA and the phase space decomposition\textemdash 
show that replacing the simplified interaction based on the OPE potential 
with the $G$-matrix in the calculation of the NN scattering amplitude results in a moderate decrease of the mUrca rate. On the other hand, the use of the nucleon propagator of Eq.~\eqref{QP:propagator} leads to a strong enhancement at all densities, and to the appearance of an 
unphysical divergence as the density approaches the activation threshold of the dUrca mechanism. 

The rate of the mUrca process was subsequently analysed by Suleiman {\it et al.}~\cite{Suleiman:2023bdf}, who carried out a detailed perturbative calculation based on the OPE interaction model, in which the treatment of thermal effects was improved by releasing the low-temperature approximations employed in Refs.~\cite{Friman:1979ecl,Shternin:2018dcn}. Higher order dynamical effects were also taken into acount by assigning a finite lifetime, $\tau$, to the quasiparticle states,
and including lowest-order vertex corrections. According to this prescription, the single-nucleon energy acquires a constant imaginary   
part, $\Gamma = \tau^{-1}$, meant to parametrise the non vanishing width of the peaks appearing in the quasiparticle spectra. The most prominent effect of the broadening of the spectral lines in the finite-temperature formalism of Ref.~\cite{Suleiman:2023bdf} turns out to be the disappearance of the divergence 
observed by Shternin {\it et al.}~\cite{Shternin:2018dcn}.
 
\subsection{Recent developments}
\label{recent}

Over the past couple of years, the dUrca and mUrca mechanisms, as well as their mutual relation, have been reexamined by the authors of Refs~\cite{Alford:2024xfb,Sedrakian:2024uma} using a relativistic formalism based on the polarisation tensor $\Pi^{\alpha \beta}$. 

Within this approach, the neutrino emissivity is written in terms of the imaginary part of the 
contraction  $L_{\alpha \beta} \Pi^{\alpha \beta}$\textemdash with $L_{\alpha \beta}$ being the lepton tensor of Eq.~\eqref{leptens}\textemdash and the contributions of 
different reaction mechanisms can be consistently taken into account by using suitable definitions of $\Pi^{\alpha \beta}$. 

\begin{center} 
\begin{figure}[h!] 
\centering
\includegraphics[width=12.00cm]{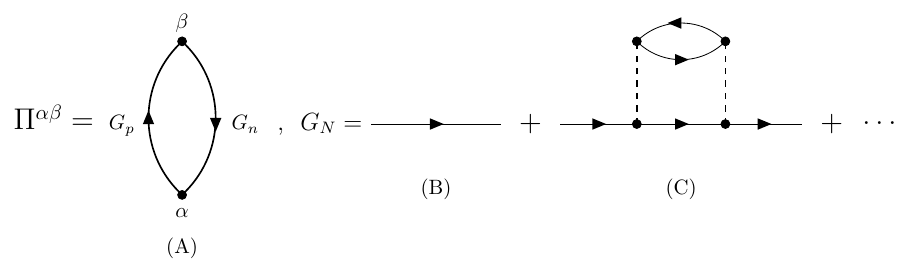}
\caption{Schematic representation of the polarisation tensor of Eq.~\eqref{poltens}, diagram (A), and lowest-order contributions to the nucleon propagator, 
diagrams (B) and (C).}
\label{pol:pict}
\end{figure}
\end{center}

In the loop approximation, schematically illustrated by the diagram of Fig.~\ref{pol:pict} (A), the polarisation tensor is defined as
\begin{align}
\label{poltens}
\Pi^{\alpha \beta} = -i \int \frac{d^4p}{(2\pi)^4} {\rm Tr} \big\{  {j^\alpha}^\dagger G_p(p+q)   j^\beta G_n(p) \big\} \ , 
\end{align}
where $j^\alpha$ and $G_N(p)$ denote the nucleon weak current and the propagator of a relativistic nucleon with four-momentum $p$,  the perturbative expansion of which is schematically 
depicted in the right panel of Fig.~\ref{pol:pict}. A comparison with the diagrams of Figs.~\ref{dUrca} and~\ref{mUrca} reveals that diagrams (B) and (C), 
correspond to the dUrca and mUrca processes, respectively. 

Within the loop approximation employed in Refs~\cite{Alford:2024xfb,Sedrakian:2024uma} the nucleon propagators include 
self-energy insertions, but vertex corrections are not taken into account. As a consequence, the Ward identities following from current conservation are not satisfied. This issue is known to be relevant in the analysis of nucleon interactions driven by the neutral weak current, the vector component of which is conserved. In the case of charged current 
interactions, on the other hand, the neutron-proton mass difference entails a violation of the Ward identities even in processes involving isolated nucleons. 

As pointed out by the authors of Ref.~\cite{Alford:2024xfb}, in nuclear matter the scale of the intrinsic violation of the Ward identities originating from isospin-symmetry breaking is dictated by the difference between the nucleon {\it effective} masses, which lies in the  tens of MeV range. At $T\lsim 10$ MeV, this difference turns out to be larger than the typical width of  quasiparticle states employed in their study.

Alford {\it et al.}~\cite{Alford:2024xfb} argued  that the distinction between dUrca and mUrca reactions is somewhat artificial, and developed a systematic approach 
in which the corresponding contributions to the emissivity can be consistently taken into account within a unified framework. In the simplest implementation of this 
scheme\textemdash dubbed Nucleon 
Width Approximation, or NWA\textemdash the propagator of a nucleon with four momrntum $k$ is written in the form
\begin{align}
\label{NWA:G}
G^{\rm NWA}_N(k,m^\star_N,\Gamma_N) =  \int_{-\infty}^{+\infty} dM~G^{\rm MFA}_N(k,M) R_N(M,\Gamma_N) \ , 
\end{align}
where $m_N^\star$ is the nucleon effective mass defined by Eq.~\eqref{effmass}, $G^{\rm MFA}_N$ denotes the relativistic generalisation of the MFA propagator of Eq.~\eqref{QP:propagator}, and $R_N(M,\Gamma_N)$ is a Breit-Wigner distribution peaked at $M = m_N^\star$, whose width parametrises the 
collisional broadening of the quasiparticle spectral lines. This procedure allows one to obtain the {\it total} neutron decay rate from the dUrca contribution using
\begin{align}
\label{NWA:rate}
\Gamma^{\rm NWA} = \int_{-\infty}^{+\infty} dM_n dM_p~\Gamma^{\rm dUrca}(M_n,M_p) R_n(M_n,\Gamma_n)  R_p(M_p,\Gamma_p) \ .
\end{align}

\begin{center} 
\begin{figure}[h!] 
\centering
\includegraphics[width=10.00cm]{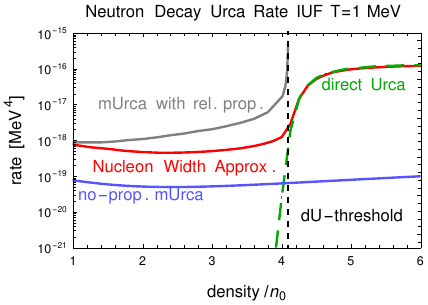}
\caption{Comparison between the density dependence of the neutron decay rate of Eq.~\eqref{NWA:rate}, and 
the results obtained using different approximations; see text for details. Reprinted from Ref.~\cite{Alford:2024xfb} with permissions. \copyright~APS 2024. All rights reserved.}
\label{alford} 
\end{figure}
\end{center}

The results of Ref.~\cite{Alford:2024xfb}, based on the so-called IUF model of the nuclear matter equation of state~\cite{IUF}, are summarised in Fig.~\ref{alford}, in which 
the density dependence of the neutron decay rate at $T=1$ MeV computed using the NWA is compared to the predictions of different 
approximation schemes. The curve labelled ``no-prop mUrca" shows the baseline results of Friman and Maxwell~\cite{Friman:1979ecl}, while the 
``mUrca with rel. prop." line has been obtained from a relativistic generalisation of the treatment of Ref.~\cite{Shternin:2018dcn}, and the long dashed 
line corresponds to the dUrca rate, computed carrying out the full phase space integration. The NWA calculation has been performed using the same nucleon width 
$\Gamma $\textemdash resulting from calculations carried out by the authors of Ref.~\cite{armen_width} using the $G$-matrix formalism\textemdash for protons and neutrons.
The emerging picture suggests a smooth transition from the low-density region, in which the emissivity is mainly driven by mUrca processes, to the 
high-density region, in which the dUrca mechanism is dominant.

In Ref.~\cite{Sedrakian:2024uma} Sedrakian reported the results of a study aimed at clarifying the impact of SRC on neutrino 
emission in the aftermath of  dUrca processes. Owing to the isospin dependence of nuclear forces, the correlated nucleon pairs 
discussed in Sec.~\ref{correlations}
predominantly consist of a proton and a neutron, the momenta of which exceed the corresponding Fermi momenta.
Electron scattering experiments have shown that in neutron-rich nuclei the deviation from the normal Fermi liquid momentum distribution 
is more pronounced for protons~\cite{hen_2014}.

Within the approach proposed by Sedrakian, in which the neutrino emissivity is obtained from the one-loop approximation to the polarisation tensor 
of Eq.~\eqref{poltens}, the non Fermi liquid behaviour of protons and neutrons is taken into account by approximating the corresponding 
spectral functions\textemdash trivially related to the Green's functions through $P_N(k) =  {\rm Im}~G_N(k)/\pi$\textemdash in the form
\begin{align}
\label{armen:spectra}
P_N(k) =  Z_k \delta \big(  k_0 - e_N(k) \big) + \frac{\Gamma_N}{ \big[ (k_0 - e_N(k)\big]^2 + \Gamma_N^2 } + \ldots \ ,      
\end{align}
obtained by expanding about $\Gamma_N=0$.
Here, the $\delta$-function term in the left-hand side corresponds to the quasiparticle approximation, whereas the second term 
takes into account the corrections arising at first order in the width $\Gamma_N$. Note that the leading contribution is corrected 
by the factor $Z_k < 1$, describing the reduced normalisation of the quasiparticle state due to SRC discussed in Sec~\ref{correlations}. 
In view of the above considerations on the isospin dependence of correlations, the broadening correction to the neutron propagator has been neglected, 
setting $\Gamma_n = 0$. On the other hand, $\Gamma_p$ has been estimated using the measured $pn$ and $pp$ scattering cross sections. 

\begin{center} 
\begin{figure}[h!] 
\centering
\includegraphics[width=8.00cm]{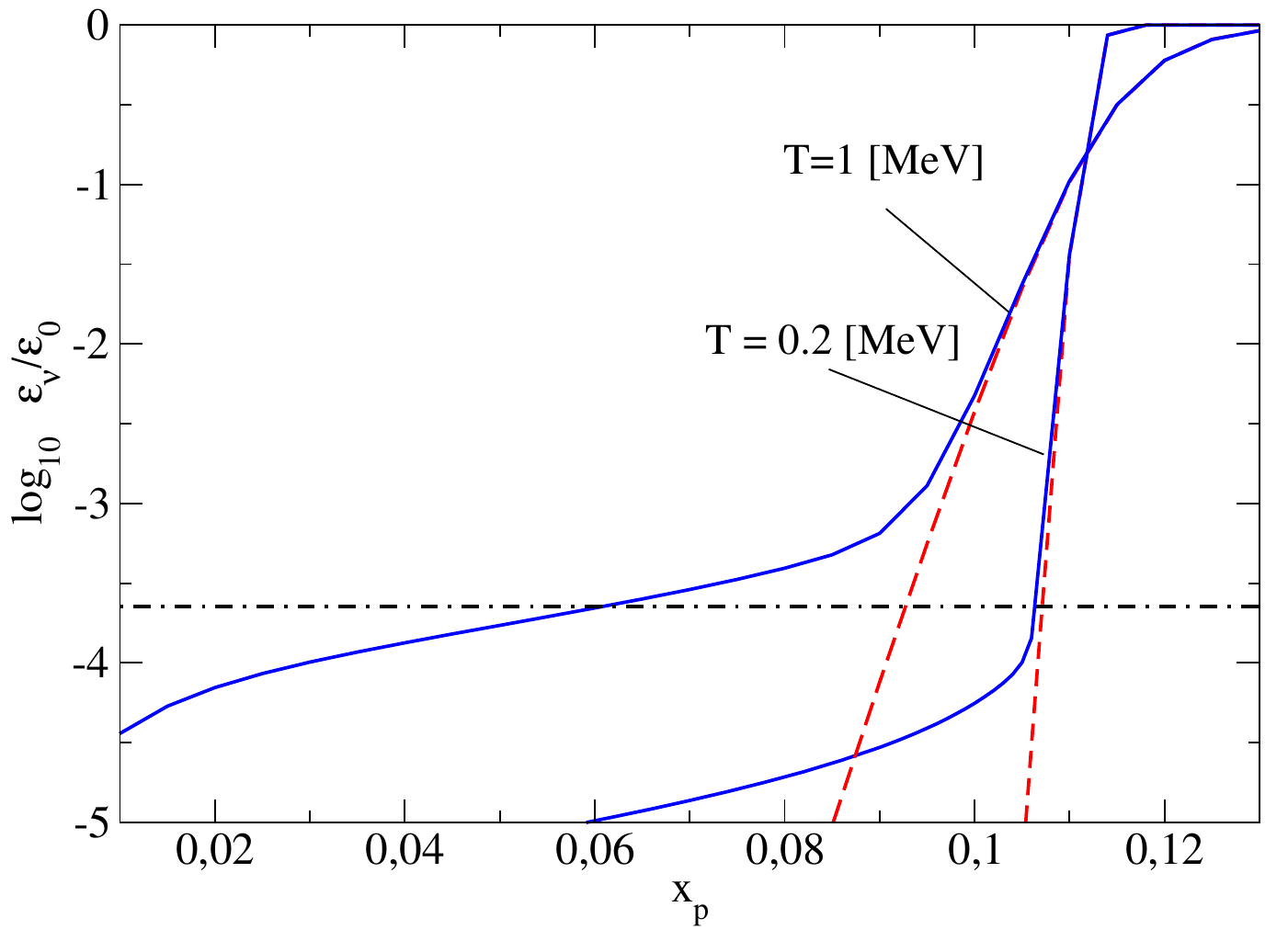}
\caption{ Proton fraction dependence of the dUrca neutrino emissivity of $npe$ matter, expressed in units of the corresponding FSA 
result. The solid and dashed lines correspond to the full calculation, performed using the spectra of 
Eq.~\eqref{armen:spectra}, and to the quasiparticle approximation, respectively. For reference, the dot-dashed horizontal line shows the mUrca 
emissivity predicted by the model of Friman and Maxwell
at $T=1$ MeV.
Reprinted from Ref.~\cite{Sedrakian:2024uma} with permissions. 
\copyright~APS 2024. All rights reserved.}
\label{armen}
\end{figure}
\end{center}

Figure~\ref{armen}, taken from Ref.~\cite{Sedrakian:2024uma}, shows the dependence of the neutrino emissivity of $npe$ matter on the proton 
fraction $x_p$, which is expected to drive the effect of SRC. The results of calculations performed using the spectra of 
Eq.~\eqref{armen:spectra}\textemdash normalised to those obtained from the FSA, Eq.\eqref{Urca:FSA}\textemdash are 
displayed for $T=$ 0.2 and 1 MeV. The solid and dashed lines correspond to the full calculation and the quasiparticle 
approximation, respectively. For reference, the mUrca emissivity predicted by the model of Friman and Maxwell
at $T=1$ MeV is represented by the horizontal dot-dashed line.
The main conclusion emerging from Fig.~\ref{armen} is that the finite quasiparticle lifetime significantly affects 
the dUrca emissivity. However, the interpretation of the observed results as correlation effects requires some qualifications.
The appearance of high-momentum nucleons, providing unambiguous evidence of SRC, reveals the presence of 
correlated nucleon pairs in the nuclear matter ground-state, which cannot be taken into account by dressing  
the propagators appearing in Eq.~\eqref{poltens} and Fig.~\ref{pol:pict} (A). The results of a detailed analysis of correlation efffects in cold SNM\textemdash carried out by the authors of Ref.\cite{PhysRevC.41.R24} using CBF perturbation theory and a realistic nuclear Hamiltonian\textemdash show that the high-momentum tail of the nucleon distribution originates from the admixture of a two-particle\textendash two-hole
component into the ground state. On the other hand, the contribution associated with self-energy insertions in the nucleon 
propagators only affects the normalisation of the quasiparticle states, $Z_k$, and turns out to be vanishing outside the Fermi sea.

\section{Summary and outlook}
\label{summary}

In recent years, a number of theoretical studies~\cite{Shternin:2018dcn,Suleiman:2023bdf,Alford:2024xfb,Sedrakian:2024uma,Lucas:Thesis} have exposed the limitations of the approximate treatment of neutrino emission 
from neutron stars matter proposed in the classic paper of Friman and Maxwell~\cite{Friman:1979ecl}, and routinely employed by many authors afterwards~\cite{Yakovlev}. 
Besides providing evidence that the assumption of full degeneracy\textemdash leading to the appearance of the activation threshold   
of the dUrca reaction\textemdash already fails at temperature as low as 1 MeV, the results of these analyses underscored the 
importance of taking into account more complex dynamical mechanisms, which lead to modifications the nucleon propagator. Moreover, they contributed to shed light 
on the connection between dUrca and mUrca processes, as well as on their role in determining the density dependence of the neutrino 
emission rate~\cite{Alford:2024xfb}.

The treatment of thermal effects has been greatly improved by carrying out accurate phase space integrations~\cite{Suleiman:2023bdf}. It must be emphasised 
that the results of these calculations are significantly affected by the temperature dependence of the proton and neutron  energies and 
chemical potentials appearing in the corresponding Fermi-Dirac distributions. Moreover, all single-nucleon properties should in principle be 
obtained from the same model of nuclear dynamics employed to determine the equation of state and composition of matter. However, these issues, thoroughly discussed in Refs.~\cite{PhysRevD.106.103020} and~\cite{Benhar:2023mgk}, are largely disregarded in the literature.

Nucleon interactions have been widely treated within the oversimplified mean-field approach, which is conceptually inadequate 
to describe the mUrca processes. On the other hand, a number of studies performed using renormalised weak current 
opeators\textemdash defined in terms of correlated wave functions obtained from realistic nuclear 
Hamiltonians\textemdash have shown that the effects of correlations are, 
in fact, significant, and lead to a sizeable quenching of the weak transition amplitudes 
driving dUrca and mUrca processes alike~\cite{cowell2003,benharfarina2009,Lovato:2012ux,lovatoetal2014}. 

It should be noted that, because the renormalised weak operator 
acts on a single nucleon, it is quite natural to interpret the quenching of the corresponding transition matrix element in terms of a vertex correction, due to the renormalisation of the nucleon wave functions. The consistent inclusion of self-energy insertions in the nucleon
   propapagtor should be analyised using the formalism of Correlated
   Basis Function (CBF) perturbation theory, as discussed in Ref.~\cite{PhysRevC.41.R24} of the manuscript. 
   Work in along this line is in progress, and the results will be reported in a forthcoming paper~\cite{tobepublished}. 

Correlation effects on the dUrca emissivity have been also studied using the one-loop approximation for the nuclear matter polarisation tensor~\cite{Sedrakian:2024uma}. However, the results of detailed calculations of the nucleon spectral functions and momentum distributions in cold SNM at equilibrium density, carried out using CBF perturbation theory and realistic nuclear Hamiltonians~\cite{FP_momdis,bff_pke,bff_Green}, show that the appearance of high-momentum nucleons, providing unambiguous evidence of correlated nucleon pairs, is associated with presence of a non-pole component in the nucleon Green's function. This contribution, which 
appears at lowest order of the CBF perturbative expansion, is not included in the model of Ref.~\cite{Sedrakian:2024uma}.  

Overall, the studies reviewed in this article, while being admittedly still limited by the lack of a consistent treatment of all quantities involved in numerical calculations, have provided valuable novel contributions, useful to shed light on the mechanisms of neutrino emission from dense and hot nuclear matter.
Based on these recent achievements, the prospects for the development of a comprehensive and reliable theoretical framework, suitable for a fully quantitative characterisation
of neutrino emissivity from neutron stars in the broad temperature and density region relevant to astronomical observations~\cite{Benhar:2024qcw}, 
appear to be promising. Attaining this goal will require the use of a microscopic model of nuclear interactions 
including both mean-field and correlation dynamics, as well as a consistent treatment of thermal effects on all nuclear matter properties at temperatures up to several tens of MeV. The approach based on non relativistic many-body theory and phenomenological nuclear Hamiltonians\textemdash which provides a quantitative description of
electron-nucleus scattering data revealing the occurrence of large density fluctuations due to NN correlations~\cite{Benhar:2006wy}\textemdash appears to have the potential to carry out this project.

\section*{Acknowledgements}
The content of this short review is partly based on a seminar given by OB at the Physics Division of Argonne
National Laboratory, the hospitality of which is gratefully acknowledged.
The authors' work has been supported by INFN\textemdash the Italian National Institute for Nuclear Research\textemdash 
under grant TEONGRAV. 

\bibliographystyle{ws-ijmpe}
\bibliography{biblio_revised}

\begin{thebibliography}{10}

\bibitem{refId0}
A.~Perego, S.~Bernuzzi and D.~Radice, {\em Eur. Phys. J. A} {\bf 55}  (2019)
  124.

\bibitem{Mezzacappa:2020oyq}
A.~Mezzacappa, E.~Endeve, O.~E. Bronson~Messer and S.~W. Bruenn, {\em Liv. Rev.
  Comput. Astrophys.} {\bf 6}  (2020)  ~4.

\bibitem{Prakash:1996xs}
M.~Prakash, I.~Bombaci, M.~Prakash, P.~J. Ellis, J.~M. Lattimer and R.~Knorren,
  {\em Phys. Rep.} {\bf 280}  (1997) 1.

\bibitem{Yakovlev:2004iq}
D.~G. Yakovlev and C.~J. Pethick, {\em Ann. Rev. Astron. Astrophys.} {\bf 42}
  (2004) 169.

\bibitem{Brandes:2023hma}
L.~Brandes, W.~Weise and N.~Kaiser, {\em Phys. Rev. D} {\bf 108}  (2023)
  094014.

\bibitem{Brandes:2024wpq}
L.~Brandes and W.~Weise, {\em Phys. Rev. D} {\bf 111}  (2025)   034005.

\bibitem{Romani:2022jhd}
R.~W. Romani, D.~Kandel, A.~V. Filippenko, T.~G. Brink and W.~Zheng, {\em
  Astrophys. J. Lett.} {\bf 934}  (2022)   L17.

\bibitem{Benhar:2023qvr}
O.~Benhar, {\em Particles} {\bf 6}  (2023) 611.

\bibitem{Yakovlev}
D.~Yakovlev, A.~Kaminker, O.~Gnedin and P.~Haensel, {\em Phys. Rep.} {\bf 354}
  (2001)  ~1.

\bibitem{Chamel:2008}
N.~Chamel and P.~Haensel, {\em Living Rev. Rel.} {\bf 11}  (2008)  ~10.

\bibitem{Friman:1979ecl}
B.~L. Friman and O.~V. Maxwell, {\em Astrophys. J.} {\bf 232}  (1979) 541.

\bibitem{Yakovlev_AA}
D.~G. {Yakovlev} and K.~P. {Levenfish}, {\em Astronomy \& Astrophysics} {\bf
  297}  (1995)   717.

\bibitem{Shternin:2018dcn}
P.~S. Shternin, M.~Baldo and P.~Haensel, {\em Phys. Lett. B} {\bf 786}  (2018)
  28.

\bibitem{Suleiman:2023bdf}
L.~Suleiman, M.~Oertel and M.~Mancini, {\em Phys. Rev. C} {\bf 108}  (2023)
  035803.

\bibitem{Alford:2024xfb}
M.~G. Alford, A.~Haber and Z.~Zhang, {\em Phys. Rev. C} {\bf 110}  (2024)
  L052801.

\bibitem{Sedrakian:2024uma}
A.~Sedrakian, {\em Phys. Rev. Lett.} {\bf 133}  (2024)   171401.

\bibitem{Lucas:Thesis}
{L. Tonetto}, Thermal effects in nuclear matter and neutron stars, PhD thesis,
  Sapienza University of Rome  (2024).
\newblock https://iris.uniroma1.it/handle/11573/1715868.

\bibitem{tobepublished}
L.~Tonetto and O.~Benhar, {\em Paper in preparation}   (2025).

\bibitem{Benhar:2024qcw}
O.~Benhar, A.~Lovato, A.~Maselli and F.~Pannarale (eds.), {\em {Nuclear Theory
  in the Age of Multimessenger Astronomy}} (CRC Press, 2024).

\bibitem{BL:2017}
O.~Benhar and A.~Lovato, {\em Phys. Rev. C} {\bf 96}  (2017)   054301.

\bibitem{Benhar_2022}
O.~Benhar, A.~Lovato and G.~Camelio, {\em Astrophys. J.} {\bf 939}  (2022)
  ~52.

\bibitem{PhysRevD.106.103020}
L.~Tonetto and O.~Benhar, {\em Phys. Rev. D} {\bf 106}  (2022)   103020.

\bibitem{Raffelt:1996wa}
G.~G. Raffelt, {\em {Stars as laboratories for fundamental physics}}
  (University of Chcago Press, 1996).

\bibitem{Hanhart:2000ae}
C.~Hanhart, D.~R. Phillips and S.~Reddy, {\em Phys. Lett. B} {\bf 499}  (2001)
  9.

\bibitem{QMC}
J.~Carlson, S.~Gandolfi, F.~Pederiva, S.~C. Pieper, R.~Schiavilla, K.~E.
  Schmidt and R.~B. Wiringa, {\em Rev. Mod. Phys.} {\bf 87}  (2015)   1067.

\bibitem{Landau}
G.~Baym and C.~J. Pethick, {\em Landau Fermi-Liquid Theory} (John Wiley \&
  Sons, 1991).

\bibitem{Benhar_NPN}
O.~Benhar, {\em Nuclear Physics News} {\bf 26}  (2016) 15.

\bibitem{RevModPhys.65.817}
O.~Benhar, V.~R. Pandharipande and S.~C. Pieper, {\em Rev. Mod. Phys.} {\bf 65}
  (Jul 1993)   817.

\bibitem{Arrington_SRC}
J.~Arrington, N.~Fomin and A.~Schmidt, {\em Annu. Rev. Nucl. Part. Sci.} {\bf
  72}  (2022) 307.

\bibitem{BF:NM}
O.~Benhar and S.~Fantoni, {\em {Nuclear Matter Theory}} (CRC Press, 2020).

\bibitem{cowell2003}
S.~T. Cowell and V.~R. Pandharipande, {\em Phys. Rev. C} {\bf 67}  (2003)
  035504.

\bibitem{CLARK197989}
J.~W. Clark, {\em Progress in Particle and Nuclear Physics} {\bf 2}  (1979)
  ~89.

\bibitem{benharvalli2007}
O.~Benhar and M.~Valli, {\em Phys. Rev. Lett.} {\bf 99}  (2007)   232501.

\bibitem{Benhar:2009xm}
O.~Benhar, A.~Polls, M.~Valli and I.~Vidana, {\em Phys. Rev. C} {\bf 81}
  (2010)   024305.

\bibitem{cowell2006}
S.~T. Cowell and V.~R. Pandharipande, {\em Phys. Rev. C} {\bf 73}  (2006)
  025801.

\bibitem{benharfarina2009}
O.~Benhar and N.~Farina, {\em Phys. Lett. B} {\bf 680}  (2009)   305.

\bibitem{Lovato:2012ux}
A.~Lovato, C.~Losa and O.~Benhar, {\em Nucl. Phys. A} {\bf 901}  (2013)  ~22.

\bibitem{lovatoetal2014}
A.~Lovato, O.~Benhar, S.~Gandolfi and C.~Losa, {\em Phys. Rev. C} {\bf 89}
  (2014)   025804.

\bibitem{Sedrakian:2018ydt}
A.~Sedrakian and J.~W. Clark, {\em Eur. Phys. J. A} {\bf 55}  (2019)   167.

\bibitem{Takatsuka:1993}
T.~Takatsuka, R.~Tamagaki and T.~Tatsumi, {\em Progress of Theoretical Physics
  Supplement} {\bf 112}  (1993) 67.

\bibitem{Pethick:RMP}
C.~J. Pethick, {\em Rev. Mod. Phys.} {\bf 64}  (1992) 1133.

\bibitem{Muto:1992}
T.~Muto, T.~Takatsuka, R.~Tamagaki and T.~Tatsumi, {\em Prog. Theor. Phys.
  Suppl.} {\bf 112}  (1993) 221.

\bibitem{Tsuruta:2002}
S.~Tsuruta, M.~A. Teter, T.~Takatsuka, T.~Tatsumi and R.~Tamagaki, {\em
  Astrophys. J. Lett.} {\bf 571}  (2002)   L143.

\bibitem{gamow:Urca}
G.~Gamow and M.~Schoenberg, {\em Phys. Rev.} {\bf 58}  (1940) 1117.

\bibitem{gamow}
G.~Gamow, {\em My World Line: An Informal Autobiography} (The Vicking Press,
  1970).

\bibitem{Benhar:LNP}
O.~Benhar, {\em {Structure and Dynamics of Compact Stars}} (Springer, 2023).

\bibitem{V6P}
R.~Wiringa and S.~Pieper, {\em Phys. Rev. Lett.} {\bf 89}  (2002)   182501.

\bibitem{UIX_2}
J.~Carlson, V.~R. Pandharipande and R.~B. Wiringa, {\em Nucl. Phys. A} {\bf
  401}  (1983)  ~59.

\bibitem{UIX}
B.~S. Pudliner, V.~R. Pandharipande, J.~Carlson and R.~B. Wiringa, {\em Phys.
  Rev. Lett.} {\bf 74}  (1995) 4396.

\bibitem{Benhar:2023mgk}
O.~Benhar, A.~Lovato and L.~Tonetto, {\em Universe} {\bf 9}  (2023)   345.

\bibitem{AV18}
R.~B. Wiringa, V.~G.~J. Stoks and R.~Schiavilla, {\em Phys. Rev. C} {\bf 51}
  (1995) 38.

\bibitem{IUF}
F.~J. Fattoyev, C.~J. Horowitz, J.~Piekarewicz and G.~Shen, {\em Phys. Rev. C}
  {\bf 82}  (2010)   055803.

\bibitem{armen_width}
{A. Sedrakian and A. E. L. Dieperink}, {\em Phys. Rev. D} {\bf 62}  (2000)
  083002.

\bibitem{hen_2014}
{O. Hen {\it et al.}}, {\em Science} {\bf 346}  (2014) 614.

\bibitem{PhysRevC.41.R24}
O.~Benhar, A.~Fabrocini and S.~Fantoni, {\em Phys. Rev. C} {\bf 41}  (1990)
  R24.

\bibitem{FP_momdis}
S.~Fantoni and V.~Pandharipande, {\em Nucl. Phys. A} {\bf 427}  (1984) 473.

\bibitem{bff_pke}
O.~Benhar, A.~Fabrocini and S.~Fantoni, {\em Nuclear Physics A} {\bf 505}
  (1989) 267.

\bibitem{bff_Green}
O.~Benhar, A.~Fabrocini and S.~Fantoni, {\em Nuclear Physics A} {\bf 550}
  (1992) 201.

\bibitem{Benhar:2006wy}
O.~Benhar, D.~Day and I.~Sick, {\em Rev. Mod. Phys.} {\bf 80}  (2008) 189.

\end{thebibliography}

\end{document}